# The wave equation and two-electron atomic wavefunctions

John Lekner

School of Chemical and Physical Sciences, Victoria University of Wellington,

Wellington, New Zealand

john.lekner@vuw.ac.nz

**Abstract** Axisymmetric solutions of the classical wave equation in three spatial dimensions are shown to be in one-to-one correspondence to functions which annihilate the two-electron kinetic energy operator $\nabla_1^2 + \nabla_2^2$ expressed in coordinates relative to a fixed nucleus.

## 1. Introduction

The Hamiltonian of the system consisting of nucleus of charge $Ze$ and two electrons is

$$H = -\frac{\hbar^2}{2m}\nabla_1^2 - \frac{\hbar^2}{2m}\nabla_2^2 - \frac{Ze^2}{r_1} - \frac{Ze^2}{r_2} + \frac{e^2}{r_{12}} \tag{1.1}$$

The motion of the nucleus is neglected. The Laplacian $L = \nabla_1^2 + \nabla_2^2$, written in terms of the distances $r_1, r_2, r_{12}$ of the electrons from the nucleus and from each other, reads [1]

$$L = \partial_1^2 + \frac{2}{r_1}\partial_1 + \partial_2^2 + \frac{2}{r_2}\partial_2 + 2\left(\partial_{12}^2 + \frac{2}{r_{12}}\partial_{12}\right) + \frac{r_1^2+r_{12}^2-r_2^2}{r_1 r_{12}}\partial_1\partial_{12} + \frac{r_2^2+r_{12}^2-r_1^2}{r_2 r_{12}}\partial_2\partial_{12} \tag{1.2}$$

Gottschalk and Maslen [2] used solutions which annihilate the kinetic energy operator to augment the Fock expansion functions in order to satisfy the physical boundary conditions. The purpose of this note is to show that axisymmetric solutions of the wave equation in ordinary-space time are in one-to-one correspondence to functions which annihilate the two-electron operator $\nabla_1^2 + \nabla_2^2$. We shall relate functions $\psi(r_1, r_2, r_{12})$ for which $L\psi = 0$ to functions $\Psi(\rho, z, t)$ satisfying the axisymmetric wave equation

$$\left\{\partial_\rho^2 + \frac{1}{\rho}\partial_\rho + \partial_z^2 - \partial_{ct}^2\right\}\Psi = 0 \qquad (\rho^2 = x^2 + y^2) \tag{1.3}$$

The proof of the relationship between these solutions will be given in terms of hyperspherical coordinates, discussed next.

## 2. Hyperspherical coordinates

The hyperspherical coordinates $r, \alpha, \theta$, are respectively the root-mean-square distance of the two electrons from the nucleus (the hyperspherical radius), an angle $\alpha$ which determines the ratio of the distances of the two electrons from the nucleus, and the angle $\theta$ subtended at the nucleus by the two electrons [3]:

$$r = \sqrt{r_1^2 + r_2^2}, \qquad \tan\frac{\alpha}{2} = \frac{r_2}{r_1}, \qquad \cos\theta = \frac{\boldsymbol{r_1}\cdot\boldsymbol{r_2}}{r_1 r_2} = \frac{r_1^2 + r_2^2 - r_{12}^2}{2 r_1 r_2} \tag{2.1}$$

The inverse relations are

$$r_1 = r\cos\frac{\alpha}{2}, \quad r_2 = r\sin\frac{\alpha}{2}, \quad r_{12} = r\sqrt{1 - \sin\alpha\cos\theta} \tag{2.2}$$

When the two electrons approach each other, $\alpha \to \frac{\pi}{2}$ and $\theta \to 0$. Then $r_{12}/r = \sqrt{1 - \sin\alpha\cos\theta} \to 0$. When one or other of the electrons approaches the nucleus, $\alpha \to 0$ or $\alpha \to \pi$. When all three charges are close together, $r \to 0$.

The cosines of the angles $\theta_1$ and $\theta_2$ at electrons 1 and 2 in the $r_1, r_2, r_{12}$ triangle, are given by

$$\cos\theta_1 = \frac{\cos\frac{\alpha}{2} - \sin\frac{\alpha}{2}\cos\theta}{\sqrt{1 - \sin\alpha\cos\theta}}, \qquad \cos\theta_2 = \frac{\sin\frac{\alpha}{2} - \cos\frac{\alpha}{2}\cos\theta}{\sqrt{1 - \sin\alpha\cos\theta}} \tag{2.3}$$

Converting the differentiations with respect to $r_1, r_2, r_{12}$ to those with respect to $r, \alpha, \theta$, we have

$$\partial_1 g = \cos\frac{\alpha}{2}\partial_r g - \frac{2}{r}\sin\frac{\alpha}{2}\partial_\alpha g + \frac{2}{r}\frac{\sin\frac{\alpha}{2}\cos\theta - \cos\frac{\alpha}{2}}{\sin\alpha\sin\theta}\partial_\theta g$$

$$\partial_2 g = \sin\frac{\alpha}{2}\partial_r g + \frac{2}{r}\cos\frac{\alpha}{2}\partial_\alpha g + \frac{2}{r}\frac{\cos\frac{\alpha}{2}\cos\theta - \sin\frac{\alpha}{2}}{\sin\alpha\sin\theta}\partial_\theta g \tag{2.4}$$

$$\partial_{12} g = \frac{2}{r}\frac{\sqrt{1 - \sin\alpha\cos\theta}}{\sin\alpha\sin\theta}\partial_\theta g$$

We note that $r_1\partial_1 + r_2\partial_2 + r_{12}\partial_{12} = r\partial_r$.

The Laplacian $L$ of (1.2) can be transformed to a Laplacian in four-dimensional space-time [3-10] by the substitutions

$$X = 2r_1 r_2 \sin\theta\cos\phi = r^2\sin\alpha\sin\theta\cos\phi$$

$$Y = 2r_1 r_2 \sin\theta\sin\phi = r^2\sin\alpha\sin\theta\sin\phi \tag{2.5}$$

$Z = 2r_1r_2\cos\theta = r^2\sin\alpha\cos\theta\,, \qquad T = r_1^2 - r_2^2 = r^2\cos\alpha$

In (2.5) the variable $r_{12}$ has been replaced by $\theta$, the angle subtended at the nucleus by the two electrons. The Laplacian is now separable: $L\psi = 0$ becomes

$$(\partial_X^2 + \partial_Y^2 + \partial_Z^2 + \partial_T^2)\psi = 0 \qquad (2.6)$$

This has the same form as the wave equation in three dimensions, in which the variable $T$ is set equal to $ict$, the phase speed multiplied by $i$ and by time. In terms of the Fock coordinates (2.5) we have

$$\rho = \sqrt{X^2+Y^2} = 2r_1r_2\sin\theta =$$
$$\sqrt{(r_1+r_2+r_{12})(r_1+r_2-r_{12})(r_1+r_{12}-r_2)(r_2+r_{12}-r_1)},$$
$$z = Z = 2r_1r_2\cos\theta = r_1^2 + r_2^2 - r_{12}^2, \qquad (2.7)$$
$$ict = T = r_1^2 - r_2^2$$

Any solution $\Psi(\rho, z, ct)$ of the axisymmetric wave equation (1.3) is thus in one-to-one correspondence with a function $\psi(r_1, r_2, r_{12})$ annihilating $L$, namely

$$\psi(r_1, r_2, r_{12}) = \Psi(2r_1r_2\sin\theta\,, 2r_1r_2\cos\theta\,, \pm i[r_1^2 - r_2^2])$$
$$= \Psi(r^2\sin\alpha\sin\theta\,, r^2\sin\alpha\cos\theta\,, \pm ir^2\cos\alpha) \qquad (2.8)$$

We have put both signs on the ‘time’ coordinate, since the wave equation is unchanged by $t \to -t$. Both signs are needed to provide $r_1, r_2$ symmetry in the quantum wavefunction.

## 3. Solutions of the wave equation, transcribed

Bateman [11] has given a general solution of the wave equation in integral form. When the wave-motion is independent of the azimuthal angle $\phi$, Bateman’s non-singular solution of $(\nabla^2 - \partial_{ct}^2)\Psi = 0$ in the coordinates $\rho, z, ct$ is

$$\Psi(\rho, z, ct) = \int_{-\pi}^{\pi} du\, F(z + i\rho\cos u\,, ct + \rho\sin u, u) \qquad (3.1)$$

This satisfies the wave equation with axial symmetry because $\nabla^2 - \partial_{ct}^2$ applied to the function $F(z + i\rho\cos u\,, ct + \rho\sin u\,, u)$ gives $-\rho^{-2}\partial_u^2 F$. Integration over $u$ from $-\pi$ to $\pi$ then gives zero. In correspondence with (3.1) we therefore have the zero kinetic energy solution

$$\psi(r_1, r_2, r_{12}) = \int_{-\pi}^{\pi} du\, F(\,2r_1 r_2(\cos\theta + i\sin\theta\cos u), \pm i(r_1^2 - r_2^2) + 2r_1 r_2 \sin\theta \sin u\,, u) \tag{3.2}$$

As a simple example, suppose that the arbitrary function $F$ takes the form $F(A, B, u) = F(A - B)$. Then, on introducing the variable $U = e^{iu}$, (3.2) reduces to

$$\psi(r_1, r_2, r_{12}) = \oint \frac{dU}{iU} F(r_1^2 + r_2^2 - r_{12}^2 \pm i(r_1^2 - r_2^2) + 2ir_1 r_2 \sin\theta\, U) \tag{3.3}$$

The integral is over the unit circle in the complex plane; the residue at $U = 0$ equals $-iF(r_1^2 + r_2^2 - r_{12}^2 \pm i(r_1^2 - r_2^2))$. Any function of $r_1^2 + r_2^2 - r_{12}^2 \pm i(r_1^2 - r_2^2)$ will therefore annihilate the Laplacian $L$, as may readily be verified. In hyperspherical coordinates

$$r_1^2 + r_2^2 - r_{12}^2 \pm i(r_1^2 - r_2^2) = r^2[\sin\alpha\cos\theta \pm i\cos\alpha] \tag{3.4}$$

The modulus of this variable is $r^2\sqrt{1 - \sin^2\alpha\sin^2\theta}$; it is zero when $r_1 = r_2$ $(\alpha = \pi/2)$ and $\theta = \pi/2$.

In addition to the Bateman solution, there are recent solutions of the wave equation in closed form, parametrized by a length $a$. Define the complex length $R(\rho, z, ct)$ by

$$R^2 = \rho^2 + (a + ict)^2 = \rho^2 - (ct - ia)^2 \tag{3.5}$$

Then an axisymmetric solution of the wave equation, for any twice differentiable function $F$, is ([12], Chapter 2 of [13], and [14])

$$\Psi(\rho, z, ct) = F(R - iz)/R \tag{3.6}$$

Further generalization is possible by a complex shift in $z$ to $F(R + b - iz)/R$. The square of the complex length $R$ and the coordinate $z$ translate respectively to

$$R^2 = 4r_1^2 r_2^2 - (r_1^2 + r_2^2 - r_{12}^2)^2 + (a^2 + r_1^2 - r_2^2)^2$$

$$z = r_1^2 + r_2^2 - r_{12}^2 \tag{3.7}$$

We have changed the space-time length $a$ to $a^2$ since the coordinates $X, Y, Z, T$ in equations (2.7) and (2.8) are of second degree in the spatial electron coordinates $r_1, r_2, r_{12}$. In hyperspherical coordinates $r, \alpha, \theta$ the relations (3.7) read

$$R^2 = r^4[1 - \sin^2\alpha\cos^2\theta] + 2r^2a^2 + a^4, \qquad z = r^2\sin\alpha\cos\theta \tag{3.8}$$

Automatic solutions of $L\psi = 0$ are therefore

$$\psi(r_1, r_2, r_{12}) = \frac{F\left(\sqrt{4r_1^2r_2^2-(r_1^2+r_2^2-r_{12}^2)^2+(a^2+r_1^2-r_2^2)^2} \pm i(r_1^2+r_2^2-r_{12}^2)\right)}{\sqrt{4r_1^2r_2^2-(r_1^2+r_2^2-r_{12}^2)^2+(a^2+r_1^2-r_2^2)^2}}$$

$$= \frac{F\left(\sqrt{r^4[1-\sin^2\alpha\cos^2\theta]+2r^2a^2\cos\alpha+a^4} \pm ir^2\sin\alpha\cos\theta\right)}{\sqrt{r^4[1-\sin^2\alpha\cos^2\theta]+2r^2a^2\cos\alpha+a^4}} \quad (3.9)$$

## 4. Existing solutions of $L\psi = 0$

Gottschalk and Maslen have obtained solutions of $L\psi = 0$ in the course of work on analytic expansions of three-body atomic wavefunctions [2]. A simple example is the logarithmic solution

$$\psi_0 = \ln\frac{1+\cos\theta}{1-\cos\theta} = \ln\frac{(r_1+r_2+r_{12})(r_1+r_2-r_{12})}{(r_1-r_2+r_{12})(r_2-r_1+r_{12})} \quad (4.1)$$

Since $\rho = 2r_1r_2\sin\theta$, $z = 2r_1r_2\cos\theta$, $\cos\theta = z/\sqrt{\rho^2+z^2} = \cos\Theta$, the polar angle in the wave equation coordinates. Indeed, $\Psi_0 = \ln\frac{1+\cos\Theta}{1-\cos\Theta}$ is a (static) solution of the wave equation.

Another logarithmic solution of $L\psi = 0$ is given in Section 3.3 of Ref. 2 as

$$\psi_1 = \ln\left[r_{12}(2r^2 - r_{12}^2)^{1/2} + r_1^2 - r_2^2\right] \quad (4.2)$$

We symmetrize this by adding to $\psi_1$ the same function with $r_1, r_2$ interchanged, to

$$\psi_s = \ln[r_{12}^2(2r^2 - r_{12}^2) - (r_1^2 - r_2^2)^2] \quad (4.3)$$

The function $\psi_s$ transcribes to $\Psi_s = \ln\rho^2$, which is again a static solution of the wave equation. Many more complicated solutions of $L\psi = 0$ are given in Ref. 2.

## 5. Discussion

Gottschalk and Maslen [2] remark in their Introduction that "Because the major part of the wavefunction is generated by the kinetic energy operator, rather than the potential, the method should also apply to non-Coulombic potentials, as well as to other atoms and molecules." By *method* is meant the use of solutions which annihilate the kinetic energy operator to augment the Fock expansion functions in order to satisfy the boundary conditions.

This paper has shown that functions which give zero when operated on by the two-electron kinetic energy are in one-to-one relation to solutions of the wave equation in ordinary space-time. The Bateman solutions of the wave equation predate modern quantum mechanics, and recent work in electrodynamics has produced another family of solutions. It is hoped that the link presented here, between classical physics and the quantum theory of the motion of two mutually repelling electrons in the attractive field of a nucleus, will be useful in solving this century-old problem.

**References**

1. E. A. HYLLERAAS, Neue Berechnung der Energie des Helium im Grundzustande, sowie des tiefsten Terms von Ortho-Helium, *Zeits. für Physik* **54** (1929) 347-366.

2. J. E. GOTTSCHALK and E. N. MASLEN, Coordinate systems and analytic expansions for three-body atomic wavefunctions III Derivative continuity via solutions to Laplace's equation, *J. Phys. A: Math. Gen.* **20** (1987) 2781-2803.

3. V. FOCK, On the Schrödinger equation of the helium atom, *Det Kongelige Norske Videnskabers Selskabs Forhandlinger* **31**, no. 22 (1958) 138-144.

4. P. C. ABBOTT and E. N. MASLEN, Coordinate systems and analytic expansions for three-body atomic wavefunctions I Partial summation for the Fock expansion in hyperspherical coordinates, *J. Phys. A: Math. Gen.* **20** (1987) 2043-2075.

5. J. E. GOTTSCHALK, P. C. ABBOTT AND E. N. MASLEN, Coordinate systems and analytic expansions for three-body atomic wavefunctions II Closed form wavefunction to second order in $r$, *J. Phys. A: Math. Gen.* **20** (1987) 2077-2104.

6. C. R. MYERS, C. J. UMRIGAR, J. P. SETHNA AND J. D. MORGAN, Fock's expansion, Kato's cusp conditions, and the exponential ansatz, *Phys. Rev. A* **44** (1991) 5537-5546.

7. E. Z. LIVERTS AND N. BARNEA, Analytic matrix elements for the two-electron atomic basis with logarithmic terms, *J. Math. Phys.* **55** (2014) 082104.

8. E. Z. LIVERTS AND N. BARNEA, Angular Fock coefficients: Refinement and further development, *Phys. Rev. A* **92** (2015) 042512.

9. E. Z. LIVERTS, Analytic calculation of the edge components of the angular Fock coefficients, *Phys. Rev. A* **94** (2016) 022504.

10. E. Z. LIVERTS AND R. KRIVEC, Averaged electron densities of the helium-like atomic systems, *J. Math. Phys*. **61** (2020) 032105.

11. H. BATEMAN, The solution of partial differential equations by means of definite integrals, *Proc. Lond. Math. Soc*. **1** (1904) 451–458.

12. J. LEKNER, Family of oscillatory electromagnetic pulses, *Phys. Rev. A* **108** (2023) 063502.

13. J. LEKNER, *Theory of electromagnetic pulses*, 2ed (IOP Publishing, Bristol, 2024).

14. J. LEKNER, Localized solutions of the wave equation, *Phys. Rev. A* **108** (2023) 063502.